\documentclass[12pt]{amsart}

\usepackage{amscd}
\usepackage{amsopn}
\usepackage{cases}

\newcommand{\Mo}{(M,\omega )}

\newtheorem{theorem}{Theorem}
\newtheorem{corollary}{Corollary}
\newtheorem{example}{Example}
\newtheorem{proposition}{Proposition}
\newtheorem{lemma}{Lemma}

\newtheorem{remark}{Remark}

\begin{document}

\title[On symplectically fat twistor bundles]{On symplectically fat twistor bundles}

\author[Boche\'nski, Szczepkowska, Tralle, Woike]{Maciej Boche\'nski$^1$, Anna Szczepkowska$^2$, Aleksy Tralle$^3$\\ and
Artur Woike$^4$}



\begin{abstract}
This paper deals with the question when are twistor bundles over homogeneous spaces  symplectically fat? It  shows that  twistor bundles over even dimensional Grassmannians of maximal rank have this property.
\end{abstract}




\maketitle

\section{Introduction}

Let  $G\rightarrow P\rightarrow B$ 
be a principal bundle with a connection.
Let $\theta$ and $\Theta$ be the connection one-form and
the curvature form of the connection, respectively.  Both forms have
values in the Lie algebra $\mathfrak{g}$ of the group $G$. 
Denote the pairing between $\mathfrak{g}$ and
its dual $\mathfrak{g}^*$ by $\langle\,,\rangle$.  By definition, a vector
$u\in\mathfrak{g}^*$ is {\bf {\em  fat}}, if the two--form
$$
(X,Y)\rightarrow \langle\Theta(X,Y),u\rangle
$$
is non-degenerate for all {\it horizontal} vectors $X,Y$. Note
that if a connection admits at least one fat vector then it admits the
whole coadjoint orbit of fat vectors. 

Let $\Mo$ be a closed symplectic manifold with a Hamiltonian action
of a Lie group $G$ and the moment map $\Psi:M\to \mathfrak g^*$. 
Consider the associated Hamiltonian bundle
$$
\Mo \to E:=P\times_G M \to B.
$$

The starting point of our research is the following result  of Sternberg and Weinstein.

\begin{theorem}[Sternberg-Weinstein]\label{thm:s-w} 
Let $\Mo$ be a a symplectic manifold with a Hamiltonian action of a
Lie group $G$ and the moment map $\Psi: M\rightarrow\mathfrak{g}^*$.
Let $G\rightarrow P\rightarrow B$ be a principal bundle. 
If there exists a connection in the principal bundle $P$ such that all
vectors in $\Psi(M)\subset\mathfrak{g}^*$ are fat, then 
 the total space of the associated bundle
$$
M\rightarrow P\times_G M\rightarrow B
$$
admits a fiberwise symplectic form.
\label{sw}
\end{theorem}
Following Weinstein \cite{wein}, we will call such associated bundles {\it symplectically fat}.
In \cite{gls} and \cite{ster} it was shown that symplectically fat fiber bundles can be used to obtain equations of motion of a classical particle in the presence of a Yang-Mills field, for any gauge group.   On the other hand, symplectically fat fiber bundles (as well as their contact analogues \cite{ler2}) are used in symplectic and contact topology to construct symplectic and contact manifolds with presribed properties. For example, the authors of \cite{ht} applied the described construction to solve problems in metric contact geometry and symplectic topology.

We see that  fat bundles are interesting and useful, both, in symplectic geometry and in mathematical physics. However, it is extremely difficult to even find  examples satisfying the fatness condition, provided that one does not assume the symplecticness of the base.  Surprisingly, the only known examples of it are the following classes of bundles (see \cite{ktw}):
\begin{enumerate}
\item bundles of the form
$$H/K\rightarrow G/K=G\times_H(H/K)\rightarrow G/H$$
where $G$ is a semisimple Lie group, $H$ is its compact subgroup of maximal rank and $K=Z_G(T)\subset H$ for some torus in $G$;
\item twistor bundles of the form:
$$SO(2n)/U(n)\rightarrow \mathcal{T}(B)\rightarrow B,$$
where $(B^{2n},g)$ stands for an even-dimensional manifold with sectional curvature $K_g$ satisfying 
the following condition
$$1- {3\over{2n+1}}\leq |K_g|\leq 1;$$
\item locally homogeneous complex  manifolds  $\Gamma\setminus G/V$ fibered over locally symmetric Riemannian manifolds as follows:
$$K/V\rightarrow \Gamma\setminus G/V\rightarrow \Gamma\setminus G/K,$$
where $G$ is a semisimple Lie group of non-compact type, $\Gamma$ is a uniform lattice in $G$, $K$ a maximal compact subgroup in $G$ and $V=Z_G(T)\subset K$ for some torus $T$ in $G$.
\end{enumerate}

 Thus, it is tempting to find some other classes of fat bundles, as well as to understand the reason why are they  so rare. 
In this work we prove that some classes of twistor bundles over homogeneous spaces are symplectically fat.
Our main result is the following theorem.
\begin{theorem}\label{thm:main} The twistor bundles over even dimensional Grassmannians of maximal rank
$$SO(2n+2m)/SO(2n)\times SO(2m),\, m,n\not=1$$
$$SO(2(n+m)+1)/SO(2n)\times SO(2m+1),n\not=1$$
$$Sp(n+m)/Sp(n)\times Sp(m)$$
$$U(m+n)/U(m)\times U(n)$$
are symplectically fat.
\end{theorem}
 Let us explain our motivation for considering twistor bundles  in more detail. In \cite{r} Reznikov proved the following. The total space of the twistor bundle over even-dimensional compact Riemannian manifold $(B^{2n},g)$ whose sectional curvature $K_g$ satisfies the inequality
$$1-{3\over {2n+1}}\leq |K_g|\leq 1,$$
admits a fiberwise symplectic structure. He used this theorem to construct examples of closed symplectic manifolds with no Kaehler structure. Note that the result of Reznikov turned out to be just a consequence of the fact that these bundles are symplectically fat. The latter result with a more conceptual proof was obtained in \cite{ktw}. 
 In this context we want also to mention an important article \cite{fp} which is related to symplectic fatness. The authors constructed there some non-Kaehler manifolds with trivial canonical bundle by using a special case  of Reznikov's theorem. All these observations lead to the natural question: {\it which twistor bundles do admit fiberwise symplectic structures?} The previous works on this topic dealt with the case when the base $(B,g)$ was endowed with a Riemannian metric satisfying some extra curvature assumptions \cite{r}, \cite{ktw}. In this article we are interested in a different type of restrictions on the base: we assume that it is a homogeneous space endowed with an invariant Riemannian metric. Our method enables us to solve the problem of symplecticness (in positive) for twistor bundles over Grassmannians. 
  Symplecticness of twistor bundles may be of independent interest, because they generalize the approach of Penrose \cite{p} which enables one to construct particular Einstein metrics on the base (see \cite{b}, Chapter 13).
Note that Theorem \ref{thm:main} is obtained as a consequence of the following approach. Analyzing the previous works on the topic, we can repeat after Weinsten \cite{wein} that ``everything in sight is homogeneous''. In this paper we approach the problem restricting ourselves to a more general but still tame case of $G$-structures over homogeneous spaces $K/H$ of semisimple Lie groups. In more detail,  

\begin{enumerate}
\item we find sufficient conditions ensuring that a $G$-structure 
$$G\rightarrow P\rightarrow K/H$$
over compact reductive homogeneous space $K/H$ admits a symplectically fat associated bundle
$$G/G_{\xi}\rightarrow P\times_{G_{\xi}}(G/G_{\xi})\rightarrow K/H$$ 
with coadjoint orbits $G/G_{\xi}$ as fibers for $\xi\in \frak g^\ast$ (Lemma \ref{lemma:fat-G-str} and Corollary \ref{cor:fat-G-str}) in terms of the isotropy representation;
\item Proposition \ref{prop:twistor-fat} (which follows from Corollary \ref{cor:fat-G-str}) and Theorem \ref{thm:semisimple}  yield conditions on the isotropy representation ensuring that the twistor bundle over even-dimensional  $K/H$ (such that $\text{rank}\,K=\text{rank}\,H$) is symplectic.
\end{enumerate}
Our main result is obtained by applying Theorem \ref{thm:semisimple} to the particular case of twistor bundles over Grassmannians. 
The constructed classes of symplectically fat fiber bundles are not covered by the previously known results. Note that the idea of using $G$-structures over homogeneous spaces is also exploited in our work \cite{bstw}. 

Finally, let us just mention that symplectic fatness condition is used in some physical models \cite{ster},\cite{hps},\cite{m1},\cite{w2}, and in the representation theory \cite{gls}. 

In what follows we use the theory of Lie groups and Lie algebras closely following \cite{dk} and \cite{ov} without further explanations.

\section{Lerman's theorem}

We need to introduce some notation which will be used throughout this work. We denote by $\mathfrak{k}$ the Lie algebra of a semisimple Lie group
$K$. The symbol $\mathfrak{k}^c$ denotes the complexification. Let $H$ be a compact subgroup of maximal rank
in $K$ with the Lie algebra $\mathfrak{h}$. Let
$\mathfrak{t}$ be a maximal abelian subalgebra in $\mathfrak{h}$. Then
$\mathfrak{t}^c$ is a Cartan subalgebra in $\mathfrak{k}^c$.  We denote by
$\Delta=\Delta(\mathfrak{k}^c,\mathfrak{t}^c)$ the system of $\mathfrak{k}^c$
with respect to $\mathfrak{t}^c$. Under these choices
the root system for $\mathfrak{h}^c$ is a subsystem of $\Delta$. Denote
this subsystem as $\Delta(\mathfrak{h})$.

If the Killing form $B_{\mathfrak{k}}$ of $\mathfrak{k}$ is nondegenerate on $\mathfrak{h}$ then the
subspace 
$$
\mathfrak{m}:= 
\{X\in \mathfrak{k}\,|\, B_{\mathfrak{k}} (X,Y)=0,\, \mbox{for all } Y\in \mathfrak{h}\,\}
$$
defines a decomposition
$$
\mathfrak{k}=\mathfrak{h}\oplus\mathfrak{m}. 
$$
The decomposition is $\operatorname{ad}_H$-invariant and the
restriction of the Killing form $B_{\mathfrak{k}}$ to $\mathfrak{m}$ is nondegenerate
(see Theorem 3.5 in Section X of \cite{kn}).
The decomposition complexifies to
$\mathfrak{k}^c=\mathfrak{h}^c\oplus\mathfrak{m}^c$. 
Thus, we have root space decompositions:
\begin{eqnarray*}
\mathfrak{k}^c
&=&\mathfrak{t}^c+\sum_{\alpha\in\Delta}\mathfrak{k}^{\alpha},\\
\mathfrak{h}^c
&=&\mathfrak{t}^c+\sum_{\alpha\in\Delta(\mathfrak{h})}\mathfrak{k}^{\alpha},\\
\mathfrak{m}^c 
&=&\sum_{\alpha\in\Delta\setminus\Delta(\mathfrak{h})}\mathfrak{k}^{\alpha}.
\end{eqnarray*}

\noindent Since $K$ is semisimple, the Killing form $B_{\mathfrak{k}}$ defines an isomorphism
$\mathfrak{k} \cong \mathfrak{k}^*$ between the Lie algebra of $K$ and
its dual. If the Killing form is nondegenerate on $\mathfrak h$,
the composition
$$
\mathfrak{h} \hookrightarrow \mathfrak{k} \stackrel{\cong}\longrightarrow 
\mathfrak{k}^* \to \mathfrak{h}^*
$$
is an $Ad_H$-equivariant isomorphism. Let us denote this isomorphism by 
$X_u\mapsto u$. Let $C\subset\mathfrak{t}$ be the Weyl chamber
and let $C_{\alpha}$ denote its wall determined by the root $\alpha$, more precisely $C_{\alpha}=\mbox {ker}\ \alpha $ is the hyperplane defined by the root $\alpha$. In what follows we will need not only the formulation of some part of Theorem \ref{T:lerman}, but also its {\it proof} in the form obtained in \cite{ktw}. Because of that, we reproduce the formulation and the proof, to make the article independent of \cite{ktw}. Note that Theorem \ref{T:lerman} was proved by Lerman \cite{ler} for compact semisimple Lie groups. Lerman's theorem expresses fatness condition in terms of some Lie algebra data related to the principal bundle

 $$H\rightarrow K\rightarrow K/H$$
 for the particular case of the canonical invariant connection. The reader can consult Section \ref{sec:G-str} where we discuss invariant connections in a more general setting of $G$-structures over homogeneous spaces.    

\begin{theorem}\label{T:lerman} 
Let $K$ be a semisimple Lie group, and $H\subset K$ a compact subgroup
of maximal rank. Suppose that the Killing form $B_{\mathfrak{k}}$ of $K$ is
non-degenerate on the Lie algebra $\mathfrak h\subset \mathfrak k$ of
the subgroup $H.$ Let $v\in \mathfrak{h}^*$. The following conditions are equivalent
\begin{enumerate}
\item
A vector $v\in\mathfrak{h}^*$ is fat with respect to the
the canonical invariant connection in the principal bundle
$$
H\rightarrow K\rightarrow K/H.
$$ 
\item
The vector $X_v$ does not belong to the set
$$
Ad_H(\cup_{\alpha\in\Delta\setminus\Delta(\mathfrak{h})}C_{\alpha}).
$$

\end{enumerate}
\end{theorem}
\noindent {\it Proof}.
The curvature form of the canonical connection in the given
principal bundle has the form

$$
\Theta(X,Y)=-\frac{1}{2}[X,Y]_{\mathfrak{h}}, \quad X,Y\in\mathfrak{m}
$$ 
(see Section \ref{sec:G-str}, Theorem \ref{thm:connection}).
Hence the fatness condition is expressed as the non-degeneracy of the
form

$$
(X,Y)\rightarrow B(X_v,[X,Y]_{\mathfrak{h}}). \quad (*)
$$
Recall that here the pairing is given by the Killing form. 
Since $X_v\in\mathfrak{h}$, $B(X_v,\mathfrak{m})=0$ and we get

$$
B(X_v, [X,Y]_{\mathfrak{h}})=B(X_v,[X,Y])=B([X_v,X],Y)
$$ 
It follows from the hypothesis that $B$ is non-degenerate on
$\mathfrak m$ and the form (*) is nondegenerate if and only if
$[X_v,X]\neq 0.$ This is equivalent to

$$
(\ker\,\operatorname{ad}_{X_v})\cap\mathfrak{m}=\{0\}.
$$
Without loss of generality we can assume that $X_v \in \mathfrak t.$
Then the last equality is, after complexification, equivalent to
the condition that 
$$
\alpha(X_v) \neq 0
$$
for all roots 
$\alpha \in \Delta \backslash \Delta(\mathfrak h)$ 
(see the root space decomposition of $\mathfrak m^c$) which means
that $X_v$ does not belong to a wall $C_{\alpha}$ 
for $\alpha \in \Delta \backslash \Delta(\mathfrak h)$. 
The general case (that is  $X_v$ is not necessarily in $\mathfrak t$)
follows since $\mathfrak{h}=\cup_{h\in H}\operatorname{Ad}_h(\mathfrak{t})$. 

\bigskip
\noindent

\hfill$\square$

\section{Symplecticness of bundles associated with $G$-structures over homogeneous spaces}\label{sec:G-str}
In this Section we will use the theory of invariant connections on homogeneous spaces in the form presented in Sections 1 and 2 of Chapter X of \cite{kn}. Let $B$ be a smooth manifold of dimension $n$, and let
$$G\rightarrow P\rightarrow B $$
be a $G$-structure, that is, a reduction of the frame bundle $L(B)\rightarrow B$ to a Lie group $G$. Any diffeomorphism $f\in \operatorname{Diff}(B)$ acts on $L(B)$ by the formula 
$$f(u):=(df_xX_1,...,df_xX_n)$$
for any frame $u=(X_1,...,X_n), X_i\in T_xB$ over a point $x\in B$. By definition, $f$ is called an automorphism of the given $G$-structure, if this action commutes with the action of $G$.

Let $B=K/H$ be a homogeneous space of a connected Lie group $K$.  Assume that $B$ is equipped with a $K$-invariant $G$-structure. The latter means that any left translation $\tau(k): K/H\rightarrow K/H$, $\tau(k)(aH)=kaH$ lifts to an automorphism. Let $o=H\in K/H$. Consider the linear isotropy representation of $H,$ that is, a homomorphism $H\rightarrow GL(T_oB)$ given by the formula
$$h \mapsto d\tau(h)_o,\,\text{for}\,h\in H,o=H\in K/H.$$ 
It is important to observe that we can fix a frame $u_o:\mathbb{R}^n\rightarrow T_oB,$ $u_{o}\in P$ and identify the linear isotropy representation of $H$ with a homomorphism $\lambda :H \rightarrow G$
$$\lambda(h)=u_o^{-1}d\tau(h)_ou_o,\,h\in H.$$
One can see this as follows. Denote by $P_{o}\subset P$ the fiber over the point $o,$ then $u_{o} \in P_{o}$.  The action of $H$ lifted to $P$ preserves $P_{o}$, hence  $h(u_{o}) \in P_{o}.$ Since the structure group $G$ acts transitively on $P_{o}$, there exists exactly one $g\in G$ such that
$$h(u_{o})=u_{o}g.$$
It is easy to see that $\lambda (h)=g.$ 

In the sequel we assume that $K/H$ is reductive. In this case $\mathfrak{k}$ can be decomposed into a direct sum
$$\mathfrak{k}=\mathfrak{h}\oplus\mathfrak{m}$$
such that $Ad_{H}(\mathfrak{m})\subset\mathfrak{m}$. Note that the latter implies $[\mathfrak{h},\mathfrak{m}]\subset\mathfrak{m}$. Also, we assume that the isotropy representation is faithful. Let us make one more straightforward but important observation.  One can  identify $\lambda$ with the restriction of the adjoint representation of $H$ on $\mathfrak{m}$  (which we also denote by $\lambda$). 

Note that the identification of the isotropy representation with the restriction of the adjoint representation on $\mathfrak{m}$ is used in \cite{kn}, Chapter X, for example, in the proof of Theorem 2.6.

We say that a connection $\theta$ in $P\rightarrow B$ is $K$-invariant, if for any $k\in K$ the lift of $\tau(k)$ preserves it. We need the following description of the set of invariant connections in the principal bundle $P\rightarrow B$ from \cite{kn}, Chapter X.

\begin{theorem} Let there be given a $K$-invariant $G$-structure over a reductive homogeneous space $B=K/H$. There is a one-to-one correspondence between the $K$-invariant connections in it, and $Ad_H$-invariant linear maps
$$\Lambda_\mathfrak{m}: \mathfrak{m}\rightarrow \mathfrak{g}.$$
\end{theorem}

\noindent Note that here $Ad_H$-invariance means that
$$\Lambda_\mathfrak{m}(Ad\,h(Z))=\lambda(h)(\Lambda_\mathfrak{m}(Z)),\,Z\in\mathfrak{m},h\in H.$$
\begin{remark} {\rm Recall that a connection in the given $K$-invariant $G$-structure is called {\it canonical} if it corresponds to the map $\Lambda_{\mathfrak{m}}=0$. }
\label{r2}
\end{remark}
The curvature of such connection is described by the following result \cite{kn1} (Theorem II.11.7).  

\begin{theorem}\label{thm:connection} The curvature form of the canonical connection in $P$ is given by the formula 
$$\Theta(X,Y)=-\lambda([X,Y]_{\mathfrak{h}}), X,Y\in\mathfrak{m}.$$
\end{theorem}
\hfill$\square$

Adopt the following notation. Let $\lambda:\mathfrak{a}\rightarrow\mathfrak{b}$ be the monomorphism of Lie algebras. Denote by $\lambda^*:\mathfrak{b}^*\rightarrow\mathfrak{a}^*$ the dual map defined by $\lambda^*(f)(X)=f(\lambda(X))$ for any $X\in \mathfrak{a}$. Let $B_{\mathfrak{b}}$ and $B_{\mathfrak{a}}$ be some non-degenerate bilinear invariant forms on $\mathfrak{b}$ and $\mathfrak{a}$ (these may be the Killing forms, for example, if the corresponding Lie algebras are semisimple). They determine the natural pairings between $\mathfrak{b}^*$ and $\mathfrak{b};$ $\mathfrak{a}^*$ and $\mathfrak{a}$. Thus
$$B_{\mathfrak{b}}(X_f,X)=\langle f,X\rangle,\,B_{\mathfrak{a}}(Y_g,Y)=\langle g,Y\rangle$$
for $f\in\mathfrak{b}^*,X\in\mathfrak{b},g\in\mathfrak{a}^*,Y\in\mathfrak{a}$. If $\lambda^*(f)\in\mathfrak{a}^*$, then the $B_{\mathfrak{a}}$-dual of $\lambda^*(f)$ will be denoted by $X^{\lambda}_f$, that is
$$B_{\mathfrak{a}}(X^{\lambda}_f,Y):=\langle\lambda^*(f),Y\rangle.$$

\begin{lemma}\label{lemma:fat-G-str} Let $K$ be a semisimple Lie group, and $H\subset K$ a compact subgroup of maximal rank. Suppose that the Killing form $K$ is non-degenerate on the Lie algebra $\mathfrak h\subset \mathfrak g$ of the subgroup $H$. Let there be given a $G$-structure over $K/H$. Assume that the isotropy representation $\lambda$ is faithful. Then $v\in\mathfrak{g}^*$ is fat with respect to the canonical connection, if the 2-form
$$B_{\mathfrak{k}}(X^{\lambda}_v,[X,Y]),\,X,Y\in\mathfrak{m}$$
is non-degenerate on $\mathfrak{m} \times \mathfrak{m}$.
\end{lemma}
\noindent {\it Proof.} By definition $v\in\mathfrak{g}^*$ is fat with respect to the canonical connection, if and only if the 2-form
$$\langle v,\Omega(X,Y)\rangle=\langle v,\lambda([X,Y]_{\mathfrak{h}})\rangle$$
is non-degenerate. Therefore
$$\langle v,\lambda([X,Y]_{\mathfrak{h}})\rangle = \langle \lambda^* v,[X,Y]_{\mathfrak{h}}\rangle=$$
$$B_{\mathfrak{h}}(X^{\lambda}_v,[X,Y]_{\mathfrak{h}})=B_{\mathfrak{k}}(X_v^{\lambda},[X,Y]_{\mathfrak{h}})=B_{\mathfrak{k}}(X^{\lambda}_v,[X,Y]).$$

\noindent Note that the last equality is proved by repeating the corresponding proof of Theorem \ref{T:lerman} in the previous section, applied to $\mathfrak{k}$.

\hfill$\square$

\begin{corollary}\label{cor:fat-G-str} Under the assumptions of the previous theorem, $v\in\mathfrak{g}^*$ is fat, if 
$$X_v^{\lambda}\not\in Ad(H)(\cup_{\alpha\in\Delta\setminus\Delta(\mathfrak{h})}C_{\alpha}).$$
\end{corollary}
\noindent {\it Proof.} Again, the argument follows by repeating verbatim the proof of Theorem \ref{T:lerman} from the previous section, since one has to prove the fatness of the vector $X^{\lambda}_v\in\mathfrak{h}\subset\mathfrak{k}$.

\hfill$\square$

\section{Symplectic fatness of twistor bundles over homogeneous spaces}

The twistor bundle over an even-dimensional Riemannian manifold $(B^{2n},g)$ is  the bundle associated with the orthonormal frame bundle of $B$ with the fiber $SO(2n)/U(n)$. In our case $B=K/H$, and our twistor bundle has the form

$$SO(2n)/U(n)\rightarrow SO(K/H)\times_{SO(2n)}(SO(2n)/U(n))\rightarrow K/H$$
where $\dim\,K/H=2n$, and $SO(K/H)$ stands for the total space of the principal $SO(n)$-bundle of oriented frames. In what follows we will denote the total space of the twistor bundle by $\mathcal{T}(K/H)$. 

 We see that if the base $B=K/H$ is homogeneous, and $g$ is $K$-invariant, the corresponding twistor bundle is associated to the $SO(2n)$-structure over $K/H.$ In this setting we can apply the results of the previous section.

In what follows we will always assume that $K$ is a semisimple Lie group, $H\subset K$ is a compact subgroup of maximal rank, and the Killing form of $K$ is non-degenerate on the Lie algebra $\mathfrak h\subset \mathfrak g$. Denote by $J$ the matrix in $\mathfrak{s}\mathfrak{o}(2n)$ consisting of $n$ blocks of the form
$$\left(
\begin{array}{cc}
0 & 1 \\
-1 & 0
\end{array}
\right) .
$$

It is known and easy to see that the homogeneous space $SO(2n)/U(n)$ is symplectic, because it is the coadjoint orbit of the dual vector $J^{*}$ (with respect to the Killing form $B_{\mathfrak{so}(2n)}$ and the standard transitive action of $SO(2n)$ on $SO(2n)/U(n)$) (section 3.4 of \cite{gls}).

\begin{proposition}\label{prop:twistor-fat} Let there be given the twistor bundle
$$SO(2n)/U(n)\rightarrow \mathcal{T}(K/H)\rightarrow K/H$$
over the reductive homogeneous space $K/H$. Let $\lambda:\mathfrak{h}\rightarrow\mathfrak{g}= \mathfrak{s}\mathfrak{o}(2n)$ be the isotropy representation. Let $J^*\in\mathfrak{s}\mathfrak{o}(2n)^*$ denote the dual to $J$ with respect to the Killing form $B_{\mathfrak{g}}$ of $\mathfrak{g}$. Assume $X_{J^*}^{\lambda}\in\mathfrak{h}$ has the  property
$$X_{J^*}^{\lambda}\not\in Ad(H)(\cup_{\alpha\in\Delta\setminus\Delta(\mathfrak{h})}C_{\alpha}).$$
Then the twistor bundle is symplecticaqlly fat.
\end{proposition} 
\noindent {\it Proof.} The proof follows from the equality
$$(X^{\lambda}_{J^*})^{\ast}=\lambda^* (J^*).$$
The latter means that $J^*$ is fat and, as a result, the coadjoint orbit of $J^*$ (or the adjoint orbit of $J$) is also fat. Thus the fiber $SO(2n)/U(n)$ has  fat image under the moment map.

\hfill$\square$

\begin{theorem}\label{thm:semisimple} Consider the twistor bundle over  reductive homogeneous space $K/H$. Assume that the following assumptions hold
\begin{enumerate}
\item $K$ is semisimple, $H$ is compact, and the Killing form of $\mathfrak{k}$ restricted to $\mathfrak{h}$ is non-degenerate;
\item $K/H$ is a reductive homogeneous space of maximal rank (that is, $\text{rank}\,K=\text{rank}\,H$);
\item there exists $T\in\mathfrak{t}\subset\mathfrak{h}$ in the Cartan subalgebra $\mathfrak{t}$ of $\mathfrak{h}$ and $\mathfrak{k}$ such that 
$$(ad\,T|_{\mathfrak{m}})^2=-id, \,T\not\in \cup_{\alpha\in\Delta\setminus\Delta(\mathfrak{h})}C_{\alpha}.$$
\end{enumerate}
Then the corresponding twistor bundle is symplectically fat.
\end{theorem}
\noindent {\it Proof.}
We prove the Theorem in two steps: first, we restrict ourselves to the case when $H$ is compact and simple, and then show how to extend the argument to the general case.

 Begin with a straightforward remark on duality: if $\lambda: V\rightarrow W$ is a monomorphism of vector spaces endowed with non-degenerate bilinear forms $B_V$ and $B_W$ such that $B_W(\lambda(u),\lambda(v))=B_V(u,v)$ then, for the duality determined by $B_V$ and $B_W$, the following holds 
$$
\text{if}\,\,  \lambda(v)=w, \,  \text{then} \, \lambda^*(w^*)=v^*. \eqno (1)
$$
Consider the case when $H$ is simple. Let $B_{\mathfrak{h}}$ be the restriction of $B_{\mathfrak{k}}$ to $\mathfrak{h}.$ It follows from the assumption that $B_{\mathfrak{h}}$ is an invariant non-degenerate form. Since $G=SO(2n)$ is a compact Lie group, the restriction $B_{\tilde{\mathfrak{h}}}$ of the Killing form $B_{\mathfrak{g}}$ of $\mathfrak{g}$ to $\tilde{\mathfrak{h}} := \lambda (\mathfrak{h})$ is also non-degenerate. Because $H$ is simple, every bilinear invariant form on $\mathfrak{h}$ is equal to the Killing form (modulo constant). Define
$$\tilde{B} (X,Y) := B_{\tilde{\mathfrak{h}}} (\lambda (X),\lambda (Y)), \ X,Y \in \mathfrak{h}.$$
Since $\lambda$ is a Lie algebra monomorphism,  $\tilde{B}$ is a non-degenerate invariant bilinear form on $\mathfrak{h}.$ We have
$$\tilde{B} = s \cdot B_{\mathfrak{h}}, \  \mbox{\rm where} \ s \neq 0.$$
Put $B_{\mathfrak{g}}:= \frac{1}{s} \cdot B_{g}.$ We obtain
$$B_{\mathfrak{g}} (\lambda (X), \lambda (Y)) = \frac{1}{s} B_{\tilde{\mathfrak{h}}} (\lambda (X), \lambda (Y)) = \frac{1}{s} s B_{\mathfrak{h}} (X,Y) = B_{\mathfrak{h}} (X,Y).$$
Thus we can apply  $(1)$ to $B_{\mathfrak{h}}$ and $B_{\mathfrak{g}}$. 

Let $\lambda(T)=ad\,T|_{\mathfrak{m}}=J$, where $J:\mathfrak{m}\rightarrow\mathfrak{m}$. Note that the latter follows from the fact, that for the canonical connection on $K/H$ the isotropy representation coincides with the adjoint representation restricted to $\mathfrak{m}$ (this is known and straightforward). It follows from  $(1)$ and our choice of $B_{\mathfrak{g}}$ that the dual $X^{\lambda}_{J^*}$ of $\lambda^*(J^*)$ must be $T$:
$$T=X_{J^*}^{\lambda}.$$
Therefore, $J^*\in\mathfrak{s}\mathfrak{o}^*(2n)$ is fat and by the assumption 3, $J^2=-id.$ Notice that $J$ is skew-symmetric with respect to the Killing form $B_{\mathfrak{k}}$. Thus $J\in\mathfrak{s}\mathfrak{o}(\mathfrak{m})$ and represents some complex structure on the vector space $\mathfrak{m}$. Since the coadjoint orbit dual to the adjoint orbit of $J$ consists of fat vectors, the proof follows.

 Consider now the general case of a compact subgroup $H$.  The proof goes essentially unchanged as in the first case, if we show the following implication
$$\mbox{if} \ \lambda (T) = J \ \mbox{then} \ \lambda^{\ast} (J^{\ast}) = T^{\ast}.$$
Without loss of generality assume that $\mathfrak{h} = \mathfrak{h}_{1} + \mathfrak{h}_{2}$ where $\mathfrak{h}_{1}, \ \mathfrak{h}_{2}$ are ideals in $\mathfrak{h}$ with trivial intersection and $\mathfrak{h}_{1}$ is a simple Lie algebra. Take $T \in \mathfrak{h}_{1}$ and assume that $\lambda (T)= J \in \mathfrak{g}.$  $B_{\mathfrak{k}}$ and $B_{\mathfrak{g}}$ Killing forms of $\mathfrak{k}$ and $\mathfrak{g},$ respectively. Take $B_{\mathfrak{h}}:= B_{\mathfrak{k}}|_{\mathfrak{h}}.$ First notice that $\mathfrak{h}_{1}$ and $\mathfrak{h}_{2}$ are $B_{\mathfrak{h}}$-orthogonal. Indeed take $H_{1} \in \mathfrak{h}_{1}, \ H_{2} \in \mathfrak{h}_{2}$. Since both spaces are ideals with trivial intersection we have $[H_{1},H_{2}]=0.$ Moreover since $\mathfrak{h}_{1}$ is simple there exist $H_{11},H_{12} \in \mathfrak{h}_{1}$ such that $[H_{11},H_{12}]=H_{1}.$ Thus
$$
B_{\mathfrak{h}}(H_{1},H_{2})=B_\mathfrak{h} ([H_{11},H_{12}],H_{2}) = B_\mathfrak{h} (H_{11},[H_{12},H_{2}])= B_\mathfrak{h} (H_{11}, 0) =0
$$
 $\tilde{\mathfrak{h}} := \lambda (\mathfrak{h})$ and set $B_{\tilde{\mathfrak{h}}} := B_{\mathfrak{g}}|_{\tilde{\mathfrak{h}}}.$ Since $\mathfrak{g}$ is a semisimple Lie algebra of compact type, $B_{\tilde{\mathfrak{h}}}$ is non-degenerate. Put $\tilde{\mathfrak{h}}_{1} := \lambda (\mathfrak{h}_{1})$ and $\tilde{\mathfrak{h}}_{2} := \lambda (\mathfrak{h}_{2}).$ Because $\lambda$ is the Lie algebra monomorphism we obtain a decomposition
$$\tilde{\mathfrak{h}} = \tilde{\mathfrak{h}}_{1} + \tilde{\mathfrak{h}_{2}},$$
where $\tilde{\mathfrak{h}}_{i}$ are ideals in $\tilde{\mathfrak{h}}$ with trivial intersection and $\tilde{\mathfrak{h}}_{1}$ is a simple Lie algebra. By a similar argument $\tilde{\mathfrak{h}}_{1}$ is $B_{\tilde{\mathfrak{h}}}$-orthogonal to $\tilde{\mathfrak{h}}_{2}.$ Let $H=H_{a}+H_{b} \in \mathfrak{h} = \mathfrak{h}_{1}+\mathfrak{h}_{2}.$ We have
$$\lambda^{\ast} (J^{\ast}) (H) = \lambda^{\ast} (J^{\ast}) (H_{a}+H_{b}) = J^{\ast} (\lambda (H_{a}+H_{b})) = B_{\mathfrak{g}} (J, \lambda (H_{a}+H_{b})) = $$
$$B_{\mathfrak{g}} (\lambda (T), \lambda (H_{a}+H_{b})) = B_{\tilde{\mathfrak{h}}} (\lambda (T), \lambda (H_{a}+H_{b}))= $$
$$B_{\tilde{\mathfrak{h}}} (\lambda (T), \lambda (H_{a})) + B_{\tilde{\mathfrak{h}}} (\lambda (T), \lambda (H_{b}))= B_{\tilde{\mathfrak{h}}} (\lambda (T), \lambda(H_{a})).$$
Since $\mathfrak{h}_{1}$ is a simple Lie algebra, we may assume (as in the first part of the proof)   that $B_{\mathfrak{h}} (X,Y)=B_{\tilde{\mathfrak{h}}} (\lambda (X), \lambda (Y))$ for any $X,Y \in \mathfrak{h}_{1}.$ Therefore one may continue as follows.
$$B_{\mathfrak{h}} (T,H_{a}) + 0= B_{\mathfrak{h}} (T,H_{a}) + B_{\mathfrak{h}} (T,H_{b}) = B_{\mathfrak{h}} (T,H) = T^{\ast} (H).$$

\hfill$\square$

\begin{corollary}\label{cor:simple} Under the assumptions of Theorem \ref{thm:semisimple}, assume that there exists an inner automorphism of $K$ of the form $Ad\,t$, $t\in H$, $t=\exp\,T$ such that 
$$(ad\,T|_{\mathfrak{m}})^2=-id, T\not\in  \cup_{\alpha\in\Delta\setminus\Delta(\mathfrak{h})}C_{\alpha}.$$
Then the twistor bundle over $K/H$ is symplectically fat.
\end{corollary}

\noindent The latter corollary yields examples of homogeneous spaces with symplectic twistor bundles over them. To describe these examples, recall that the compact real form of any semisimple complex Lie algebra $\mathfrak{g}^c$ can be written using the following formula
$$\mathfrak{g}=\sum_{\alpha\in\Delta}\mathbb{R}(iH_{\alpha})+\sum_{\alpha\in\Delta}\mathbb{R}(X_{\alpha}-X_{-\alpha})+\sum_{\alpha\in\Delta}\mathbb{R}(i(X_{\alpha}+X_{-\alpha})).$$
Here $\Delta$ denotes the root system for $\mathfrak{g}^c$.

This observation enables us to compute examples of bundles described in Theorem \ref{thm:semisimple}. Assume that we are given a compact homogeneous space $K/H.$ Denote by $\mathfrak{k},\mathfrak{h}$ Lie algebras of $K,H$ and by $\mathfrak{k}^c, \ \mathfrak{h}^c$ complexifications of these algebras. Since $\mathfrak{k}$ is of compact type,  the restriction of its Killing form to $\mathfrak{h}$ is non-degenerate and $K/H$ is  reductive. We also obtain the following decompositions
$$\mathfrak{k}^c=\mathfrak{t}^c+\sum_{\alpha\in\Delta(\mathfrak{h})}\mathfrak{k}_{\alpha}+\sum_{\beta\in\Delta\setminus\Delta(\mathfrak{h})}\mathfrak{k}_{\beta}$$
$$\mathfrak{h}^c=\mathfrak{t}^c+\sum_{\alpha\in\Delta(\mathfrak{h})}\mathfrak{k}_{\alpha},$$
$$\mathfrak{m}^c=\sum_{\beta\in\Delta\setminus\Delta(\mathfrak{h})}\mathfrak{k}_{\beta}.$$
Here $\Delta$ again denotes the root system for $\mathfrak{k}$. Assume that we can choose $T\in\mathfrak{t}^c$ satisfying the equations 
$$\alpha(T)\in i\mathbb{R},\,\alpha\in\Delta(\mathfrak{h})\,\mbox{and}\,\alpha(T)=\pm i,\,\alpha\in\Delta\setminus\Delta(\mathfrak{h}),$$
where $i$ or $-i$ are chosen in a way to ensure that the above system of linear equations has a solution.
Note that  $\mathfrak{t}^c=\mathfrak{t}+i\mathfrak{t}$, where $\mathfrak{t}$ denotes the real form of $\mathfrak{t}^c$ consisting of vectors $H\in\mathfrak{t}^c$ such that $\alpha(H)\in\mathbb{R}$ for all $\alpha\in\Delta$. It follows that $T\in i\mathfrak{t}=\sum_{\alpha\in\Delta}\mathbb{R}(iH_{\alpha})$. Therefore, $T\in\mathfrak{k}$. But $(ad\,T|_{\mathfrak{m}})^2=-id$, because by construction it satisfies this equality on $\mathfrak{m}^c$. Clearly, $T$ does not belong to any wall $C_{\alpha},\alpha\in\Delta\setminus\Delta(\mathfrak{h})$. Finally, we see that under the adopted assumptions the twistor bundle over $K/H$ must be symplectically fat.

\begin{example}\label{ex:f4} The twistor bundle
$$SO(16)/U(8)\rightarrow\mathcal{T}(F_{4}/SO(9))\rightarrow F_{4}/SO(9), $$
over the Riemannian symmetric space $F_{4}/SO(9)$ is symplectically fat.
\end{example}
\noindent {\it Proof.}
It is sufficient to find an appropriate $T\in\mathfrak{t}^c.$ We have
$$\Delta = \left\{  \pm e_{s} \pm e_{t}, \ \pm e_{s}, \ \frac{\pm e_{1} \pm e_{2} \pm e_{3} \pm e_{4}}{2} \ | \ 1\leq s,t \leq 4 \right\}$$ 
$$\Delta (\mathfrak{h}) = \{ \pm e_{s} \pm e_{t}, \ \pm e_{s} \  | \ 1\leq s,t \leq 4  \},$$
therefore
$$\Delta \backslash \Delta (\mathfrak{h}) = \left\{ \frac{\pm e_{1} \pm e_{2} \pm e_{3} \pm e_{4}}{2} \right\}.$$
Take $T\in\mathfrak{t}^c$ so that $e_{1}(T)=2i$ and $e_{2}(T)=e_{3}(T)=e_{4}(T)=0.$ Then   $\alpha(T)=\pm i,\,\alpha\in\Delta\setminus\Delta(\mathfrak{h})$ and
$$\alpha(T)\in i\mathbb{R},\,\alpha\in\Delta(\mathfrak{h}).$$
\hfill$\square$

\noindent
In the same fashion one can analyze the second example.

\begin{example}\label{ex:g2} The twistor bundle
$$SO(6)/U(3)\rightarrow\mathcal{T}(G_{2}/SU(3))\rightarrow G_{2}/SU(3),$$
 is symplectically fat.
\end{example}

\section{Twistor bundles over Grassmannians}

In this Section we will prove Theorem \ref{thm:main}.
 
 We will examine twistor bundles over oriented Grassmannian homogeneous spaces
$$SO(n)/SO(n-k)\times SO(k),$$
$$SU(n)/S(U(n-k)\times U(k)),$$
$$Sp(n)/Sp(n-k)\times Sp(k).$$
Since in this article we treat homogeneous spaces of maximal rank, and of even dimension,  our attention is limited to the following cases:
$$SO(2n+1)/SO(2n+1-k)\times SO(k), \ k \neq 2,$$
$$SO(2(n+m))/SO(2n)\times SO(2m), \ n,m\neq 1,$$
$$Sp(n)/Sp(n-k)\times Sp(k).$$
\vskip6pt
\centerline{\bf Proof of Theorem \ref{thm:main}}
\vskip6pt
The proof is obtained by applying Theorem \ref{thm:semisimple} to each case of even dimensional Grassmannians of maximal rank separately.
\begin{enumerate}
\item  The twistor bundle
$$SO(4nm)/U(2nm)\rightarrow\mathcal{T}(SO(2n+2m)/SO(2n)\times SO(2m))$$
$$\rightarrow SO(2n+2m)/SO(2n)\times SO(2m), \ n,m\neq 1$$
over the Riemannian symmetric space 
$$SO(2n+2m)/SO(2n)\times SO(2m)$$ 
is symplectically fat.
To prove this, 
we need to choose  $T\in\mathfrak{t}^c \cap \mathfrak{so}(2m)$ satisfying  the assumptions of Theorem \ref{thm:semisimple}. We have
$$\Delta = \{  \pm e_{s} \pm e_{t}, \ | \ 1\leq s,t \leq n+m \}$$ $$\Delta (\mathfrak{h}) = \{ \pm e_{s} \pm e_{t}  \ | \ 1\leq s,t \leq n\} \cup \{ \pm e_{s} \pm e_{t}\ | \ n+1\leq s, t\leq n+m  \}.$$
Take $T\in \mathfrak{so}(2m)$ so that
$$
e_s(T)=\begin{cases} 0, &\text{for}\,\, 1\leq s\leq n\\
i , &\text{for}\,\, n+1\leq s\leq n+m.
\end{cases}
$$
Since any root $\alpha\in\Delta\setminus\Delta(\mathfrak{h})$ is of the form $\pm e_{s} \pm e_{t}$ $s\leq n;$  $t>n,$ thus $\alpha(T)=\pm e_{t} (T) = \pm i$ and
$$\alpha(T)\in i\mathbb{R},\,\alpha\in\Delta(\mathfrak{h}).$$

\item The twistor bundle
$$SO(4nm+2n)/U(2nm+n)\rightarrow\mathcal{T}(SO(2(n+m)+1)/SO(2n)\times SO(2m+1))$$
$$\rightarrow SO(2(n+m)+1)/SO(2n)\times SO(2m+1),  n\neq 1,$$
over the Riemannian symmetric space 
 $$SO(2(n+m)+1)/SO(2n)\times SO(2m+1)$$ 
 is symplectically fat.
Again, we choose an appropriate $T\in\mathfrak{t}^c \cap \mathfrak{so}(2n).$ We have
$$\Delta = \{  \pm e_{s} \pm e_{t}, \pm e_{s} \ | \ 1\leq s,t \leq n+m \}$$ $$\Delta (\mathfrak{h}) = \{ \pm e_{s} \pm e_{t}  \ | \ 1\leq s,t \leq n\} \cup \{ \pm e_{s} \pm e_{t}, \pm e_{s}\ | \ n+1\leq s, t\leq n+m  \}.$$
Take $T\in \mathfrak{so}(2n)$ so that
$$
e_s(T)=\begin{cases} 
i & \textrm{ for }  1\leq s\leq n,\\
0 & \textrm{ for }  n+1\leq s\leq n+m.
\end{cases}
$$
Since any root $\alpha\in\Delta\setminus\Delta(\mathfrak{h})$ is of the form $\pm e_{s} \pm e_{t}$ or $\pm e_{s}$ $s\leq n;$ $t>n,$ thus $\alpha(T)=\pm e_{s} (T) = \pm i$ and
$$\alpha(T)\in i\mathbb{R},\,\alpha\in\Delta(\mathfrak{h}).$$
If $m=0$ then it is sufficient to take $e_{s} (T) = i$ for $1\leq s \leq n.$

\item The twistor bundle
$$SO(4nm)/U(2nm)\rightarrow\mathcal{T}(Sp(n+m)/Sp(n)\times Sp(m))$$
$$\rightarrow Sp(n+m)/Sp(n)\times Sp(m),  $$
over the Riemannian symmetric space 
$$Sp(n+m)/Sp(n)\times Sp(m)$$ 
is symplectically fat.

We argue as in the previous cases.  We have
$$\Delta = \{  \pm e_{s} \pm e_{t}, \pm 2e_{s} \ | \ 1\leq s,t \leq n+m \}$$ $$\Delta (\mathfrak{h}) = \{ \pm e_{s} \pm e_{t}, \pm 2e_s  \ | \ 1\leq s,t \leq n\} \cup \{ \pm e_{s} \pm e_{t}, \pm 2e_s \ | \ n+1\leq s, t\leq n+m  \}.$$
Take $T\in \mathfrak{sp}(m)$ so that
$$
e_s(T)=\begin{cases}
0, & \textrm{ for }  1\leq s\leq n,\\
i, &  \textrm{ for }  n+1\leq s\leq m.
\end{cases}
$$ 
Since any root $\alpha\in\Delta\setminus\Delta(\mathfrak{h})$ is of the form $\pm e_{s} \pm e_{t}$ $s\leq n;$  $t>n,$ thus $\alpha(T)=\pm e_{t} (T) = \pm i$ and
$$\alpha(T)\in i\mathbb{R},\,\alpha\in\Delta(\mathfrak{h}).$$
\item The case of complex Grassmannian does not require a separate proof, because $U(n+m)/U(m)\times U(n)$ is Kaehler, hence, symplectic, and the proof follows from \cite{wein}, Theorem 3.3.

\end{enumerate}

\noindent Theorem \ref{thm:main} is proved. 

\hfill$\square$

\end{document}